# Acoustic instability of a circular vortex with a smoothed vorticity profile

*V. F. Kopiev, S. A. Chernyshev, A. B. Barbasov*

*Central Aerohydrodynamic Institute (TsAGI)*

*Moscow, 105005 Russia*

e-mail: vkopiev@mktsagi.ru

It is known that a localized vortex can have two specific mechanisms of interaction with the ambient flow. The first mechanism is associated with acoustic radiation, which is accompanied by a loss of energy and causes instability in the case of negative energy of vortex disturbances. The second is a Miles mechanism of interaction of the vortex core oscillations with disturbances in the vicinity of the critical layer (where the phase velocity of the disturbances coincides with the velocity of the mean flow), accompanied by an energy flux from the critical layer vicinity, which leads to damping in the case of negative energy of the oscillations. For the first time, the flow with both of these mechanisms is considered. The problem is solved from the first principles. It is shown that the Miles mechanism can completely suppress acoustic instability, however, in the case of a stronger loss of energy due to acoustic radiation, acoustic instability will dominate. The role of various parameters is analyzed and a quantitative criterion for the acoustic instability of a vortex with a smoothed vorticity profile is obtained.

**Introduction**

The problem of acoustic instability of vortex flows is considered. The peculiarity of this problem is that vortex disturbances can have negative energy (the energy of the perturbed state is less than the energy of the undisturbed flow), which can lead to excitation of the disturbances with loss of energy for radiation, and, accordingly, to instability of the flow. For the first time, the effect of acoustic instability was obtained in [1], where a 2-dimensional circular Kelvin vortex with a stepped vorticity profile was considered as the mean flow. An explanation of the physical causes of acoustic instability was given in [2].

Of interest is an universality of the obtained solution and, in particular, the question of the sensitivity of the acoustic instability to a change in the vorticity profile. Thus, in [3] it was stated that vortex oscillations of the discrete spectrum with negative energy in the case of energy loss due to acoustic radiation are unstable. In [4] it was shown that perturbations of vortices with non-increasing vorticity in a unbounded flow have negative energy and the statement was made that all such flows are unstable. At the same time, the question of the existence of discrete spectrum in a problem discussed in [3] was not considered. Later, in [5, 6] it was shown that the acoustic instability of plane vortex flows with circular streamlines is a rather subtle effect. Under certain conditions, there is no the discrete spectrum is absent, and then instability may not occur. In particular, for the case of smoothed vorticity profile, the flow is stable in the limit of small compressibility [5].

However, as shown in this work, such a flow can become unstable with increasing the energy losses due to radiation. For this purpose, the stability of a flow with a circular vortex is considered for the case when two factors (vorticity profile smoothness and compressibility) are comparable in order of magnitude. It is shown that under certain conditions the discrete spectrum mode disappears and acoustic instability does not appear. Otherwise, the modes of the discrete spectrum always turn out to be acoustically unstable.

In contrast to the stepped vortex, in the case of a smoothed vorticity profile, qualitatively new effects arise associated with the energy flux from the vicinity of the critical layer into perturbations of the rest flow [7] (the so-called Miles interaction mechanism [8]). In the case of incompressible flow, this effect leads to the damping of the discrete mode. In this case the flow perturbations have a continuous spectrum, and the discrete eigen-frequency moves to the non-physical sheet of the complex frequency plane, and the corresponding eigen-oscillation transform to the so-called "quasi-mode" [9]. Thus, on the one hand, compressibility leads to an increase of the imagine part of eigen-frequency (increment) and on the other hand, the presence of a critical layer leads to damping of the oscillations and a decrease of the increment. As a result, the flow can be stable or unstable depending on the ratio of the flow parameters.

The article also considers a problem in which similar mechanisms of energy exchange between the oscillations of the vortex core and the surrounding flow are realized in the case of an incompressible fluid. In this problem, a vortex having a smoothed vorticity profile is located in a region bounded by an absorbing boundary, providing an alternative way of energy loss. In [10], a similar problem was considered for the Rankine vortex, which turns out to be unstable due to the energy flux through the impedance wall. In this paper, we consider a more complex flow with monotonically decreasing vorticity, which provides competition between two mechanisms that have an opposite effect on the stability of the flow.

It should be noted that consideration of the subtle effects discussed in the work makes sense only for those flows that, in the absence of these effects, are hydrodynamically stable. Otherwise, hydrodynamic instability will destroy the structure of the flow regardless of the influence of additional factors. Therefore, the work considers only vortices with a monotonically decreasing (non-increasing) vorticity profile. As noted above, such a flow in an infinite fluid is a state with maximum energy for all iso-vortex disturbances, and therefore is hydrodynamically stable.

In the first section, the oscillations of a circular vortex are considered in a limited region with an impedance boundary in the incompressible fluid approximation. The flow includes a vortex core with constant vorticity, surrounded by a small, monotonically decreasing vorticity. In this case, there is a discrete spectrum of oscillations, and the factor leading to instability is the energy flux to the external boundary of the domain. In the second section, an unbounded flow of compressible gas is considered. The solution obtained in the first section is used to analyze the instability of the system depending on the Mach number and the structure of the vorticity field in the vicinity of the critical layer. The boundary of the instability and the magnitude of the increment are obtained.

The result of the work is of interest from the point of view of understanding the process of sound generation by vortex perturbations, as well as the inverse effect of emitted sound waves on the dynamics of vorticity.

### 1. Incompressible flow with a circular vortex in a limited area

This section considers linear perturbations of the 2-dimensional flow of an incompressible ideal fluid in a bounded circular domain $\rho < R$, where $\rho$ – the radial coordinate. The mean flow has vorticity $\Omega(\rho)$ with a profile depending on the radial coordinate. It is assumed that the vorticity is constant $\Omega(\rho) = \Omega_0$ in the region $\rho < a$, and in the area $a < \rho < R$ vorticity decreases monotonically from the value $\Omega(\rho) = \Omega_1$ to zero (Fig.1). The value of vorticity in the region $a < \rho < R$ is assumed to be small, that is, $\Omega_1/\Omega_0 = O(\varepsilon)$, where $\varepsilon \ll 1$.

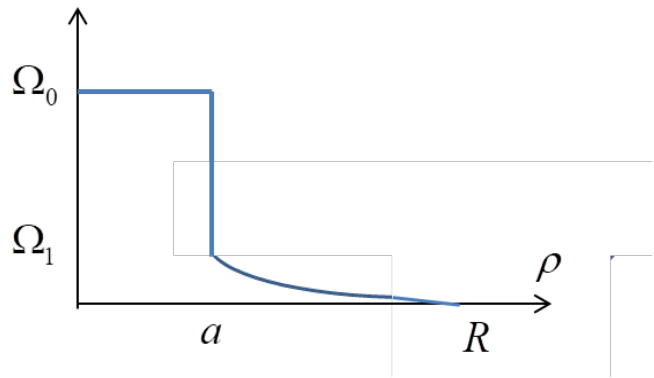

Fig.1. Vorticity profile of the mean flow.

Linear harmonic perturbations of such a flow of the form $\exp(-i\omega t + im\varphi)$, where ω – frequency, φ – angular coordinate, are considered. Such perturbations are described by the equation for the stream function *A* [7]:

$$\frac{d^2 A}{d\rho^2} + \frac{1}{\rho}\frac{dA}{d\rho} - \left(\frac{m^2}{\rho^2} + \frac{m\Omega'(\rho)}{\rho(mU(\rho) - \omega)}\right) A = 0 \qquad (1)$$

where *U* – the angular velocity of the mean flow, the components of the velocity perturbations are expressed in terms of the stream function by the equations:

$$v^\rho = \frac{imA}{\rho}, \qquad v^\varphi = -\frac{dA}{d\rho} \qquad (2)$$

The boundary value problem for eigen-frequencies is solved for equation (1) with the condition that the solution is bounded at $\rho = 0$ and the condition at the external boundary $\rho = R$

$$-\frac{R}{m}\frac{A'}{A} = 1 + \mu \qquad (3)$$

Taking into account (2) and the equation $p = \frac{\rho}{m}(\omega - mU)A'$, where the gas density is taken to be equal to unity, condition (3) is an impedance-type condition relating the normal component of the velocity and the pressure. In particular, at $\mu = -1$ condition (3) corresponds to $p = 0$, that is, a soft boundary, while at $\mu = \infty$ this condition corresponds to $v^\rho = 0$, that is, a hard boundary, and at $\mu = 0$ this condition corresponds to a "transparent" boundary/ In the last case a solution coincide with the solution in an unbounded flow. The latter follows from the fact that in the case of unbounded flow the stream function will have the form $A = const \rho^{-m}$ and, respectively, $-\frac{\rho}{m}\frac{A'}{A} = 1$. In this work we will assume that $\text{Im}\,\mu > 0$. As will be shown below, this corresponds to the absorption of energy by the outer boundary $\rho = R$. Also, we assume that $\mu$ is a small value $\mu = O(\varepsilon)$.

First of all, consider the perturbations in the region $\rho < a$. In this region, the vorticity is constant and equation (1) has the form

$$\frac{d^2 A}{d\rho^2} + \frac{1}{\rho}\frac{dA}{d\rho} - \frac{m^2}{\rho^2}A = 0 \tag{4}$$

The solution to this equation with the condition of boundedness at $\rho = 0$ is the following

$$A(\rho) = \left(\frac{\rho}{a}\right)^m \tag{5}$$

At the boundary of the vortex core $\rho = a$ the radial component of the velocity perturbations and, accordingly, the stream function is continuous, and the discontinuity of the derivative of the stream function can be expressed through the derivative of the mean vorticity field at this boundary, $\Omega'(\rho) = (\Omega_1 - \Omega_0)\delta(\rho - a)$, where $\delta(x)$ is delta function. Substituting this expression into equation (1), we obtain in the neighborhood of the point $\rho = a$

$$\frac{d^2 A}{d\rho^2} = \frac{m(\Omega_1 - \Omega_0)}{\rho(mU(a) - \omega)} A\delta(\rho - a) \tag{6}$$

where $U(a) = \frac{\Omega_0}{2}$. From (6) we get a discontinuity in the derivative of A at $\rho = a$

$$[A'] = -\frac{m(\Omega_1 - \Omega_0)}{\rho\left(m\frac{\Omega_0}{2} - \omega\right)} A \tag{7}$$

From (5) and (7) we obtain the ratio of the stream function and its derivative on the external side of the vortex core boundary

$$\left.\frac{A'}{A}\right|_{\rho=a+0} = \frac{m}{a} - \frac{m}{a}\frac{\Omega_0 - \Omega_1}{m\frac{\Omega_0}{2} - \omega} \qquad (8)$$

The equation (8) allows us to replace the original boundary value problem in the area $0 < \rho < R$ to the task in the area $a < \rho < R$ with boundary conditions (3) and (8).

Following [7], we write a particular solution to equation (1) in the form

$$A_1 = \left(\frac{\rho}{a}\right)^{-m}\left[1 + \frac{1}{2m}\left(\left(\frac{\rho}{\rho_c}\right)^{2m} - 1\right)\varepsilon g(\rho_c)\rho_c \ln(\rho - \rho_c)(1 + \varepsilon q(\rho)) + \varepsilon h(\rho)\right] \qquad (9)$$

where the critical layer $\rho = \rho_c$ is determined by the coincidence of the phase velocity of disturbances with the mean velocity, that is $\frac{\omega}{m} = U(\rho_c)$, $\varepsilon g(\rho_c) = \frac{\Omega'(\rho_c)}{\rho_c U'(\rho_c)}$,

$h(\rho) = \sum_{k=2}^{\infty} h_k(\rho - \rho_c)^k$, $q(\rho) = \sum_{k=1}^{\infty} q_k(\rho - \rho_c)^k$. Note that the logarithmic singularity in solution (9) for $\rho = \rho_c$ appears due to the presence of a regular singular point in the coefficient of the equation (1) [11]. The second linearly independent partial solution $A_2(\rho)$ of the equation (1) is obtained from (9) by replacing $m \to -m$.

We look for the solution of equation (1) with boundary conditions (3) and (8) in the form

$$A = A_1 + cA_2 \qquad (10)$$

where c - is a complex constant of order of magnitude $c = O(\varepsilon)$. From (9) and (10) we obtain

$$\frac{A'}{A} = -\frac{m}{\rho} + \left(\frac{\rho}{\rho_c}\right)^{2m}\frac{1}{\rho}\varepsilon g(\rho_c)\rho_c \ln(\rho - \rho_c) +$$
$$+ \frac{1}{2m}\left(\left(\frac{\rho}{\rho_c}\right)^{2m} - 1\right)\frac{\varepsilon g(\rho_c)}{(\rho - \rho_c)}\rho_c + \varepsilon h'(\rho) + \frac{m}{\rho}\left(\frac{\rho}{a}\right)^{2m}c + O(\varepsilon^2) \qquad (11)$$

For a Kelvin vortex, which has a stepped vorticity profile, the eigen-frequency is $\omega_0 = (m-1)\frac{\Omega_0}{2}$, and the critical layer is located on the streamline $\rho = \rho_0$, where $\rho_0 = a\sqrt{\frac{m}{m-1}}$. In the problem under consideration, we represent the eigen-frequency in the form $\omega = \omega_0 + \delta\omega$, where $\delta\omega$ - complex quantity, the imaginary part of which for unstable oscillations $\mathrm{Im}\,\delta\omega > 0$. Then the location point of the critical layer is also complex and has the form $\rho_c = \rho_0 + \delta\rho$, where $\mathrm{Im}\,\delta\rho = \frac{\mathrm{Im}\,\delta\omega}{mU'_0(\rho_0)} + O(\varepsilon^2)$. Because of $U'_0(\rho_0) < 0$, then for unstable oscillations $\mathrm{Im}\,\delta\rho < 0$. Taking into account the smallness of the value $\delta\rho = O(\varepsilon)$ write the logarithmic term in (11) for real values $\rho$ in the form of

$\ln(\rho - \rho_c) = \ln|\rho - \rho_0| + i\pi\theta(\rho_0 - \rho)$, where θ is the step function. Then, neglecting the terms of the order $O(\varepsilon^2)$ rewrite (11) in the form

$$\frac{A'}{A} = -\frac{m}{\rho} + \left(\frac{\rho}{\rho_0}\right)^{2m} \frac{\rho_0}{\rho} \alpha \left(\ln|\rho - \rho_0| + i\pi\theta(\rho_0 - \rho)\right) +$$

$$+ \frac{1}{2m}\left(\left(\frac{\rho}{\rho_0}\right)^{2m} - 1\right)\frac{\alpha}{(\rho - \rho_0)}\rho_0 + \varepsilon h'(\rho) + \frac{m}{\rho}\left(\frac{\rho}{a}\right)^{2m} c + O(\varepsilon^2) \quad (12)$$

where $\alpha = \dfrac{\Omega'(\rho_0)}{\rho_0 U_0'(\rho_0)}$. Take the imaginary part of the equation (12)

$$\mathrm{Im}\frac{A'}{A} = \left(\frac{\rho}{\rho_0}\right)^{2m}\frac{\rho_0}{\rho}\alpha\pi\theta(\rho_0 - \rho) + \frac{m}{\rho}\left(\frac{\rho}{a}\right)^{2m}\mathrm{Im}\,c + O(\varepsilon^2) \quad (13)$$

Substitute expression (13) in boundary conditions (3) and (8). For $\rho = a$ we obtain

$$\left(\frac{a}{\rho_0}\right)^{2m}\frac{\rho_0}{a}\alpha\pi + \frac{m}{a}\mathrm{Im}\,c = -\frac{m}{a}4\frac{\mathrm{Im}\,\delta\omega}{\Omega_0} + O(\varepsilon^2) \quad (14)$$

For $\rho = R$ we obtain

$$\frac{m}{R}\left(\frac{R}{a}\right)^{2m}\mathrm{Im}\,c = -\frac{m}{R}\mathrm{Im}\,\mu + O(\varepsilon^2) \quad (15)$$

From (14) and (15) we obtain the increment value

$$\mathrm{Im}\frac{\delta\omega}{\Omega_0} = -\pi\alpha\frac{\rho_0}{4m}\left(\frac{a}{\rho_0}\right)^{2m} + \frac{1}{4}\left(\frac{a}{R}\right)^{2m}\mathrm{Im}\,\mu + O(\varepsilon^2) \quad (16)$$

From (16) it follows that $\mathrm{Im}\,\delta\omega > 0$, that is, the system is unstable under the condition

$$\mathrm{Im}\,\mu > \pi\alpha\frac{\rho_0}{m}\left(\frac{R}{\rho_0}\right)^{2m} \quad (17)$$

If condition (17) is not satisfied, then the eigen-frequency moves to the non-physical sheet of the complex plane and the discrete mode disappears [9].

Let us consider this problem from the point of view of the balance of oscillation energy. We use expressions for the density and flux of perturbations energy obtained in [12]. Writing these expressions for the case of a circular vortex in an incompressible fluid, we obtain the angle-averaged values of the disturbance energy density $\langle E \rangle$ and radial component of energy flow $\langle S^\rho \rangle$

$$\langle E \rangle = -\frac{1}{2}\frac{m}{\rho}\frac{\omega_R \Omega}{(\omega_R - mU)^2 + \omega_I^2} AA^* \mathrm{Re}\left(\frac{A'}{A} + \frac{mU'}{\omega - mU}\right) \quad (18)$$

$$\langle S^\rho \rangle = -\frac{1}{2} \omega_R A A^* \operatorname{Im}\left( \frac{A'}{A} + \frac{m}{\rho} \frac{\Omega}{\omega - mU} \right) \qquad (19)$$

where $\omega_R = \operatorname{Re}\omega$, $\omega_I = \operatorname{Im}\omega$, the density of the gas is taken equal to unity. Integrating (18) over the vortex core region, $\rho < a$, find the energy of disturbances in this region

$$E = 2\pi \int_0^a \langle E \rangle \rho d\rho = -\pi m(m-1) + O(\varepsilon) \qquad (20)$$

Taking into account the normalization constant, this expression coincides with the results obtained in [2, 13]. Note that the perturbation energy (20) is negative, and in the case of an energy flux from the vortex core, the oscillation amplitude will increase over time. Substituting (11) into (19), we write down the total energy flux $S = 2\pi\rho \langle S^\rho \rangle$ in the region outside the vortex core $a < \rho < R$ in the form of

$$S = -\pi\omega_R \left(\frac{a}{\rho_0}\right)^{2m} \alpha\rho_0 \operatorname{Im}\ln(\rho - \rho_c) + \pi\omega_R m \left(\frac{a}{R}\right)^{2m} \operatorname{Im}\mu +$$

$$+ \pi\omega_R \left(\frac{a}{\rho}\right)^{2m} m \frac{\Omega\omega_I}{(\omega_R - mU)^2 + \omega_I^2} + O(\varepsilon^2) \qquad (21)$$

The first term in this equation has a logarithmic singularity at the point $\rho = \rho_c$, which provides a break in the energy flux when passing through the vicinity of the critical layer. The third term makes a significant contribution to (21) in the vicinity of the critical layer $\rho_0 - \Delta < \rho < \rho_0 + \Delta$, where $\Delta/\rho_0 = O(\sqrt{\varepsilon})$. Outside this vicinity this term can be neglected. Thus, from (21) we obtain the energy flux in the region $a < \rho < \rho_0 - \Delta$, directed from the core to the critical layer,

$$S_1 = -\alpha\pi^2 \omega_R \rho_0 \left(\frac{a}{\rho_0}\right)^{2m} + \pi\omega_R m \left(\frac{a}{R}\right)^{2m} \operatorname{Im}\mu + O(\varepsilon^2) \qquad (22)$$

and energy flux in the region $\rho_0 + \Delta < \rho < R$, directed from the critical layer to the external boundary,

$$S_2 = \pi\omega_R m \left(\frac{a}{R}\right)^{2m} \operatorname{Im}\mu + O(\varepsilon^2) \qquad (23)$$

We assume that $\operatorname{Im}\mu > 0$. In this case, the energy flux to the outer impedance boundary is positive, which corresponds to the absorption of energy at this boundary.

The magnitude of the increment is determined by the ratio of the energy flux from the vortex core region and the perturbation energy in this region, that is $\operatorname{Im}\delta\omega = -S_1/2E$. Substituting expressions (20) and (22) into this equation, we obtain the increment value that coincides with (16).

## 2. A circular vortex in an unlimited area, taking into account compressibility

This section considers an unbounded vortex flow, which has vorticity profile defined above in the region $\rho < R$, and potential flow in the area $\rho > R$. The problem is considered in the weak compressibility approximation. In this case, the factor causing instability of the system is not the non-reflective boundary at $\rho = R$, but the compressibility effect. In this case, the factor causing instability of the system will be not the energy-absorbing boundary at $\rho = R$, as in the case of Section 1, but the radiation of the energy with sound waves.

Taking into account the small Mach number of the flow, the length of the acoustic waves will be much larger than the size of the vortex and in the region $\rho < R$ the solution obtained above in the incompressible fluid approximation can be used [14]. On the stream line $\rho = R$ this inner solution is stitched with the outer solution in the region $\rho > R$, which represents an outgoing acoustic wave.

First of all, consider the perturbations in the region $\rho > R$. Taking into account the smallness of the Mach number, we neglect the convection of disturbances and write down the velocity potential for a sound wave with the condition of radiation at infinity through the Hankel function

$$\Phi(\rho) = H_m^{(1)}(k\rho) \tag{24}$$

Limiting ourselves further to the case of the second harmonic, $m = 2$, we write down the decomposition (24) in the near region, $k\rho \ll 1$,

$$\Phi = -\frac{i}{\pi}\left(\left(\frac{k\rho}{2}\right)^{-2} + 1 - \left(\frac{k\rho}{2}\right)^2 \left(\ln\frac{k\rho}{2} + C - \frac{3}{4} - i\pi\frac{1}{2}\right)\right) + O(k^4 \rho^4) \tag{25}$$

From (25) we obtain at $\rho = R$

$$iR\frac{v^\varphi}{v^\rho} = -\frac{2}{R}\frac{\Phi}{\Phi'} = 1 + \frac{k^2 R^2}{4} - \frac{k^4 R^4}{16}\left(2\ln\frac{kR}{2} + 2C - 1 - i\pi\right) + O(k^6 R^6) \tag{26}$$

Comparing (26) and (3), we obtain that the outgoing sound wave gives the same ratio of the velocity components for $\rho = R$, as the boundary condition (3) with the value $\mu$

$$\mu = \frac{k^2 R^2}{4} - \frac{k^4 R^4}{16}\left(2\ln\frac{kR}{2} + 2C - 1 - i\pi\right) \tag{27}$$

Substituting (27) into (16), we obtain the increment value

$$\text{Im}\frac{\omega}{\Omega_0} = -\pi\alpha\frac{\rho_0}{8}\left(\frac{a}{\rho_0}\right)^4 + \pi\frac{k^4 a^4}{64} + O(\varepsilon^2) \tag{28}$$

Using the expression $\alpha = \dfrac{\Omega'(\rho_0)}{\rho_0 U_0'(\rho_0)}$, we obtain from (28), that $\text{Im}\,\omega > 0$, that is, the system is unstable, under the condition

$$k^4 a^4 > 4\rho_0 \frac{|\Omega'(\rho_0)|}{\Omega_0} \tag{29}$$

If there is no vorticity derivative in the critical layer, then from (28) follows the well-known result of acoustic instability [2] with increment $\pi \frac{k^4 a^4}{64}$. When decreasing vorticity appears in the critical layer, the increment decreases, but if the Miles damping is small compared to the compressibility effect, then the instability of the discrete mode remains. At increase of the vorticity derivative, the increment decreases and at the value $|\Omega'(\rho_0)| = \frac{\Omega_0 k^4 a^4}{4\rho_0}$ the eigen-frequency reaches the real axis of the frequency complex plane and crosses the cut corresponding to the continuous spectrum of the system. At further increase of the vorticity derivative, the mode disappears from the top sheet of the Riemann surface and moves under the cut.

Now, we will consider this problem from the point of view of energy balance. In this system, just like in the system considered in Section 1, there is an energy flux from the region $a < \rho < R$, which in this case is caused not by the impedance wall, but the radiation of the sound wave. We determine the magnitude of this energy flow by substituting the value $\mu$ from (27) in (23),

$$S_2 = \pi^2 \omega_R \frac{k^4 a^4}{8} + O(\varepsilon^2) \tag{30}$$

Accordingly, the energy flux from the region of the vortex core is the following

$$S_1 = -\alpha \pi^2 \omega_R \rho_0 \left(\frac{a}{\rho_0}\right)^4 + \pi^2 \omega_R \frac{k^4 a^4}{8} + O(\varepsilon^2) \tag{31}$$

Using expressions (20) for the energy of disturbances in the vortex core and energy flux (31), we obtain the increment value $\operatorname{Im} \delta\omega = -\frac{S_1}{2E}$, coinciding with (28).

The instability of oscillations is due to the negative sign of the perturbation energy $E < 0$ in the core of the vortex and the positive sign of the energy flux $S_1 > 0$ from the core. If the energy loss to the acoustic radiation is equal to the amount of energy coming from the critical layer vicinity, then the instability disappears. Thus, the analysis of the energy balance allows an alternative way to determine the magnitude of the increment and the limits of the instability of a compressible flow with a circular vortex with a smoothed vorticity profile.

### Discussion

As is known, neutrally stable perturbations of the discrete spectrum of the vortex flow turn out to be unstable in the presence of energy-absorbing walls or energy losses due to acoustic radiation (the so-called radiation instability of the vortex). It is shown in this work that the Miles mechanism, which provides the positive energy flux to the vortex core from the vicinity of the critical layer, reduces the increment of the radiation instability. With an increase in Miles damping proportional to the derivative of vorticity in the critical layer, the discrete mode moves from the upper sheet of the Riemann frequency surface to the non-physical sheet through a cut associated with a continuous spectrum of disturbances, and it leads to the disappearance of instability.

In the work, a criterion is obtained that determines the boundaries of instability depending on the parameters of the flow. This criterion is consistent with the statement made in [5] that an arbitrarily small compressibility cannot make a vortex flow acoustically unstable. However, as can be seen from (28), with increasing compressibility, the instability of a vortex with a smoothed vorticity profile turns out to be quite realizable. It should also be noted that the damping of acoustic instability associated with the energy flux from the vicinity of the critical layer is realized only in the case when the derivative of the mean vorticity in the critical layer is nonzero. In particular, if a vortex has a smoothed vorticity profile, but in the vicinity of the critical layer there is a region with constant vorticity, then there is no energy flux from the vicinity of the critical layer, and such a vortex will be acoustically unstable at arbitrarily low compressibility.

**Funding**

The work was carried out within the framework of the RF State Assignment, 18.01.2023, № 020-00006-23-00

**References**

1. E. G. Broadbent, D. W. Moore. Acoustic destabilization of vortices, Phil. Trans. R. Soc. 1979. V. A 290. P. 353-371.
2. V. F. Kop'ev, E. A. Leont'ev, On acoustic instability of axial vortex // Akust. Zh. 1983. V. 29, №2. P. 192-198.
3. V. F. Kop'ev, E. A. Leont'ev, Some remarks to Lighthill's theory in connection with, sound radiation by compact vortexes // Akust. Zh. 1986. V. 32, №2. P. 184-189.
4. V. F. Kop'ev and E. A. Leont'ev, Acoustic instability of plane vortex flows with circular streamlines // Akust. Zh. 1988. V. 34, №3. P. 475-480.
5. S.D. Danilov. On acoustic instability of a flow with circular streamlines // Akust. Zh. 1989. V. 35, №6. P. 1059-1065.
6. R. J. Briggs, J. D. Daugherty, R. H. Levy. Role of Landau Damping in Crossed Field Electron Beams and Inviscid Shear Flow // Physics of Fluids.1970. V. 13, №2. P. 421-432
7. V. F. Kopiev, S. A. Chernyshev, Instability of an Oscillating Cylinder in a Circulation Flow of Ideal Fluid // Fluid Dynamics. 2000. V. 35, №6. P. 858-871.
8. J.W. Miles, On the generation of surface waves by shear flows // J. Fluid Mech., 1957. V.3, , №2, P. 185-204
9. D. A. Schecter, D. H. E. Dubin, A. C. Cass, C. F. Driscoll, I. M. Lansky, and T. M. O'Neil. Inviscid damping of asymmetries on a two-dimensional vortex // Physics of Fluids. 2000. V. 12, №10. P. 2397-2412.
10. E. G. Broadbent, D. W. Moore. The Two-Dimensional Instability of an Incompressible Vortex in a Tube with Energy-Absorbent Walls, Proc. R. Soc. Lond. A. 1994. V. 446. P. 39-56.
11. Olver F.W.J. Introduction to Asymptotics and Special Functions. Academic Press. 1974. 297 P.

12. Kopiev V.F., Chernyshev S.A. On Using Lagrangian Mechanics Methods to Analyze the Energy Balance in Compressed Gas Vortex Flows // Akust. Zh. 2021. V. 67, №1. P. 98-106.
13. V. F. Kopiev, S. A. Chernyshev, Vortex ring oscillations, the development of turbulence in vortex rings and generation of sound // Russian Acad. of Sciences, Phisics-Uspehi. 2000. V. 43, №7. P. 663-690.
14. M.J. Lighthill. On sound generated aerodynamically: I. general theory // Proc. Royal Soc. Series A. 1952. V. 211. P. 564–581.